\documentclass[3p,times,twocolumn]{elsarticle}

\usepackage{ecrc}

\usepackage{psfrag} % LaTeX

\volume{00}

\firstpage{1}

\journalname{Nuclear Physics B Proceedings Supplement}

\runauth{}

\jid{nuphbp}

\jnltitlelogo{Nucl.Phys.B\\Proc.Suppl.}

\usepackage{amssymb}
\newcommand{\bean}{\begin{eqnarray*}}
\newcommand{\eean}{\end{eqnarray*}}

\newcommand{\gapproxeq}{\lower
.7ex\hbox{$\;\stackrel{\textstyle >}{\sim}\;$}}
\newcommand{\lapproxeq}{\lower
.7ex\hbox{$\;\stackrel{\textstyle <}{\sim}\;$}}

\newcommand\lsim{\mathrel{\rlap{\lower4pt\hbox{\hskip1pt$\sim$}}
    \raise1pt\hbox{$<$}}}
\newcommand\gsim{\mathrel{\rlap{\lower4pt\hbox{\hskip1pt$\sim$}}
    \raise1pt\hbox{$>$}}}
\newcommand{\ba}{\begin{array}}
\newcommand{\ea}{\end{array}}
\newcommand{\nn}{\nonumber}

\newcommand{\be}{\begin{equation}}
\newcommand{\ee}{\end{equation}}
\newcommand{\bear}{\begin{eqnarray}}
\newcommand{\eear}{\end{eqnarray}}

\newcommand{\cO}{{\cal O}}

\newcommand{\mL}{\mathcal{L}}

\newcommand{\mF}{\mathcal{F}}

\newcommand{\mM}{\mathcal{M}}

\newcommand{\Frac}[2]{\frac{\displaystyle #1}{\displaystyle #2}}
\newcommand{\Int}{\displaystyle{\int}}
\begin{document}

\begin{frontmatter}

%%%  REQUESTED ON 29-X-2014
%%%\preprint{IFT-UAM/CSIC-14-113}
%%%\preprint{FTUAM-14-45}



\dochead{}
%% e.g. \dochead{17th International Conference on Dynamical Processes in Excited States of Solids}

\title{
\vspace*{-7cm} 
\\
\hspace*{13cm} 
{\small IFT-UAM/CSIC-14-113}   
\\
\hspace*{14cm}  
{\small FTUAM-14-45}   
\\
\vspace*{5cm} 
Electroweak chiral Lagrangian with a light Higgs \\
and $\gamma\gamma\to Z_L Z_L, W_L^+ W_L^-$  scattering at one loop}


%\author{}

%\address{}

\author{R.L. Delgado, A. Dobado,}
%\Address{\affiliation1}
\address{
Departamento de F\'isica Te\'orica I, UCM, \\
Universidad Complutense de Madrid,
  Avda. Complutense s/n,
28040 Madrid, Spain}

\author{M.J. Herrero and  J.J. Sanz-Cillero 
}
%\Address{\affiliation2}

\address{
Departamento de F\'isica Te\'orica and Instituto de F\'isica Te\'orica,
IFT-UAM/CSIC\\
Universidad Aut\'onoma de Madrid,  C/ Nicol\'as Cabrera 13-15, \\
Cantoblanco, 28049 Madrid, Spain
%%%\vspace*{-0.505cm}
}

\begin{abstract}
In these proceedings we provide a brief summary of the findings of a previous article
where we have studied the
photon-photon scattering into longitudinal weak bosons within the context of the electroweak chiral
Lagrangian with a light Higgs, a low-energy effective field theory including a Higgs-like scalar singlet
and where the electroweak would-be
Goldstone bosons are non-linearly realized. We consider the relevant Lagrangian up to next-to-leading order
in the chiral counting, which is explained in some detail here. We find that these amplitudes
are ultraviolet finite and the relevant combinations of next-to-leading parameters  ($c_\gamma$ and
$a_1-a_2+a_3$) do not get renormalized. We propose the joined analysis of $\gamma\gamma$--scattering
and other photon related observables ($\Gamma(h\to\gamma\gamma)$, $S$--parameter and the
$\gamma^*\to W^+_LW^-_L$ and $\gamma^*\to h\gamma$ electromagnetic form-factors) in order to
separate and determine each chiral parameter. Moreover, the correlations  between observables
provided by the NLO computations would lead to more stringent bounds on the new physics that
is parametrized by means of this effective Lagrangian.
We also show an explicit computation of the $\gamma\gamma$--scattering up to next-to-leading order
in the $SO(5)/SO(4)$ minimally composite Higgs model.

\end{abstract}


\end{frontmatter}

\vspace*{-0.0cm}
\section{$\gamma \gamma$--scattering as a probe into new physics}

%\vspace*{-0.35cm}

Two years ago the Large Hadron Collider (LHC) discovered a new
particle, most likely a scalar,  with mass $m_h\approx 125$~GeV~\cite{LHC-exp}
and couplings so far compatible with what one would expect for the Standard Model (SM) Higgs boson.
We are therefore in a scenario with small deviations from the SM and, apparently,
a large mass gap (as no new particle has shown up below the TeV).
Thus, the effective field theory (EFT) framework seems to    be  the     most    convenient    one     to      confront    current
experimental \  \ data  \  \ and  \ \  to  \ explore \  \  possible \  beyond
\\
{}\\
Standard Model (BSM) effects in the electroweak (EW) sector.

In these proceedings we discuss some of the findings
in a previous work~\cite{photon-scat}.
Therein we studied the processes
$\gamma\gamma\to Z_L Z_L$ and $\gamma\gamma\to W_L^+ W_L^-$
in the context of a general EW low-energy effective field theory (EFT),
which we will denote as Electroweak Chiral Lagrangian with a light Higgs (ECLh),
with the EW would-be Goldstone bosons (WBGBs) denoted here by $w^a$ and non-linearly realized.
In addition to be more general, this non-linear representation seems to be more appropriate
in the case of strong interactions in the EW sector, as it is the case
in Quantum Chromodynamics~\cite{Weinberg:1979,chpt}.
The three would-be Goldstone bosons $w^a$ from the spontaneous EW symmetry breaking
are parametrized through a unitary matrix $U$ that
takes values in the $SU(2)_L\times SU(2)_R/ SU(2)_{L+R}$ coset.~\footnote{
Two parametrizations of the coset were considered in Ref.~\cite{photon-scat}:
exponential coordinates, $U=\exp\{ i \tau^a w^a /{\rm v}\}$;
and spherical coordinates, $U=\sqrt{1-w^a w^a/{\rm v}^2} + i \tau^a w^a/{\rm v}$.  Both parametrizations
are found to produce the same prediction for the $\gamma\gamma\to w^a w^b$ amplitudes
once the external particles are set on-shell. Other representations of $U$ were recently studied
in~\cite{Machado:2014}.
}

The Higgs boson does not enter in the SM at tree-level in these $\gamma\gamma\to V_L V_L$  ($V=Z,W$) processes
(where in addition ${  \mM(\gamma\gamma\to Z_LZ_L)^{\rm tree}_{\rm SM}=0  }$~\cite{Jikia:1993}). Nevertheless, one can search
for new physics by studying the one-loop corrections~\cite{photon-scat},
which are sensitive to deviations from the SM in the Higgs boson couplings.
Our analysis~\cite{photon-scat} is performed in the Landau gauge and making use of the
Equivalence Theorem (Eq.Th.)~\cite{equivalence-theorem},
\bear
\mM(\gamma\gamma\to W_L^a W_L^b) &\simeq& \, -\, \mM(\gamma\gamma\to w^a w^b) \, ,
\eear
valid in the energy regime $m_{W}^2, m_{Z}^2\ll s$.
The EW gauge boson masses $m_{W,Z}$ are then neglected
in our computation.
Furthermore, since experimentally $m_h\sim m_{W,Z}$
we also neglect $m_h$ in our calculation.
In summary, the applicability range in~\cite{photon-scat} is
\bear
m_h^2 \, \sim\, m_W^2,\, m_Z^2  \quad \stackrel{\rm Eq.Th.}{\ll} \quad
s,\, t,\, u \quad \stackrel{\rm EFT}{\ll}  \quad \Lambda_{\rm ECLh}^2 \, ,
\eear
with the upper limit given by the  EFT cut-off  $\Lambda_{\rm ECLh}$,
expected to be of the order  of $4\pi {\rm v}\simeq 3$~TeV or the mass of possible heavy BSM particles,
where ${\rm v}=246$~GeV denotes the SM Higgs vacuum expectation value.
%%%At the practical level, the $W,Z,h$ masses were neglected in our analysis.

Although our derivation is general and does not assume any particular underlying BSM theory,
it is obviously inspired by models where the Higgs
is another (pseudo) Nambu-Goldstone boson (NGB).
Indeed, in the final part of these proceedings we provide an explicit example for the $SO(5)/SO(4)$ Minimally Composite
Higgs Model (MCHM)~\cite{MCHM}.

\section{ECLh up to next-to-leading order}

The WBGBs are described by a matrix field $U$ that takes values
in the $SU(2)_L \times SU(2)_R/SU(2)_{L+R}$ coset,   and transforms as
$U \to L U R^\dagger$~\cite{Appelquist:1980vg,Longhitano:1980iz}.
The basic building blocks employed to construct the relevant ECLh Lagrangian for
our analysis are~\cite{photon-scat,Appelquist:1980vg,Longhitano:1980iz}
\begin{eqnarray}
%%%U(w^\pm,z) &=& 1 + i w^a \tau^a/{\rm v} + \cO(w^2)\;\in SU(2)_L \times SU(2)_R/SU(2)_{L+R},
%%%\label{Umatrix}\\
%%%{\cal F}(h)&=& 1+2a\frac{h}{{\rm v}}+b \left(\frac{h}{{\rm v}} \right)^2+\dots ,
%%%\label{polynomial}\\
&&
\hspace*{-1.5cm}
D_\mu U =\partial_\mu U + i\hat{W}_\mu U - i U\hat{B}_\mu \, ,
\;
%%%\nn\\
%%%&&
V_\mu = (D_\mu U) U^\dagger \, ,\;
\label{VmuandT}
\label{eq.cov-deriv}
\nn \\
&&
\hspace*{-1.5cm}
\hat{W}_{\mu\nu} = \partial_\mu \hat{W}_\nu - \partial_\nu \hat{W}_\mu
 + i  [\hat{W}_\mu,\hat{W}_\nu ],
\quad %\;
\hat{B}_{\mu\nu}  = \partial_\mu \hat{B}_\nu -\partial_\nu \hat{B}_\mu ,
\label{fieldstrength}
\nn\\
&&
\hat{W}_\mu = g W_\mu^a  \tau^a/2 ,\;\hat{B}_\mu = g'\, B_\mu \tau^3/2 \, .
\label{EWfields}
\end{eqnarray}
with well-defined transformation properties~\cite{photon-scat,Longhitano:1980iz}.
The Higgs field $h$ is a singlet in the ECLh and enters in the Lagrangian operators via polynomials or
their partial derivatives~\cite{ECLh-Op4,Pich:2013}.
These building blocks are employed to construct ECLh operators with CP,
Lorentz and $SU(2)_L \times U(1)_Y$ gauge invariance.

We consider the following scaling in powers of momentum $p$,
\bear
\partial_\mu\, ,m_W\, , m_Z, \,  m_h \sim \cO(p) \, ,   \quad
g,\, g', \, e \sim \cO(p/{\rm v})\, ,
\eear
and the counting for the tensors above~\cite{photon-scat,EW-chiral-counting,chpt+photons},
\bear
D_\mu U,\; V_\mu\sim \cO(p),\quad
\hat{W}_{\mu\nu},\;\hat{B}_{\mu\nu} \sim  \cO(p^2)
\,.
\eear

Within the approximations of our analysis~\cite{photon-scat}, the relevant ECLh operators
for $\gamma \gamma \to w^a w^b$ at leading order (LO) --$\cO(p^2)$--
and next-to-leading order (NLO) in the chiral counting --$\cO(p^4)$--
are~\cite{photon-scat,Longhitano:1980iz}
\bear
&& \hspace*{-1.25cm}
\mL_2 =    -\Frac{1}{2 g^2} {\rm Tr} ( \hat{W}_{\mu\nu}\hat{W}^{\mu\nu})
-\Frac{1}{2 g^{'2}} {\rm Tr} ( \hat{B}_{\mu\nu} \hat{B}^{\mu\nu})
\nn\\
&&\hspace*{-1.25cm}
+\Frac{{\rm v}^2}{4}\left[%
  1 + 2a \Frac{h}{{\rm v}} + b \Frac{h^2}{{\rm v}^2}\right] {\rm Tr} ( D^\mu U^\dagger D_\mu U )
 + \Frac{1}{2} \partial^\mu h \, \partial_\mu h
 %%%-\frac{1}{2}m_h^2 h^2
 + \dots\, ,
%% \nn\\
%%\label{eq.L2}
\nn \\
&& \hspace*{-1.25cm}
\mL_4 =
  a_1 {\rm Tr}(U \hat{B}_{\mu\nu} U^\dagger \hat{W}^{\mu\nu})
  + i a_2 {\rm Tr} (U \hat{B}_{\mu\nu} U^\dagger [V^\mu, V^\nu ])
\nn\\
&& \hspace*{-1.25cm}
  - i a_3  {\rm Tr} (\hat{W}_{\mu\nu}[V^\mu, V^\nu])
 -\Frac{c_{\gamma}}{2}\Frac{h}{{\rm v}}e^2 A_{\mu\nu} A^{\mu\nu}\, +\, ...
%%%
%%%  - c_{W} \Frac{h}{{\rm v}} {\rm Tr}(\hat{W}_{\mu\nu} \hat{W}^{\mu\nu})
%%%  - c_B \Frac{h}{{\rm v}}\, {\rm Tr} (\hat{B}_{\mu\nu} \hat{B}^{\mu\nu} ) + \dots
\label{eq.L4}
\label{eq.ECLh}
\eear
where  one has the photon field strength  $A_{\mu \nu}=\partial_\mu A_\nu- \partial_\nu A_\mu$
and the dots stand for operators not relevant
within our approximations for the $\gamma\gamma$-scattering~\cite{photon-scat}.

The classification of the chiral order in the previous Lagrangian~(\ref{eq.ECLh})
provides a consistent perturbative expansion as we show now in more detail.
First, we denote as $\cO(p^d)$ any operator of the generic form
\bear
\mL_d &=& \sum_k f_k^{(d)} p^d \left(\frac{\chi}{{\rm v}}\right)^k\, ,
\label{eq.Ld}
\eear
with $\chi$ any bosonic field  ($h$, $w^a$, $W^a_\mu$, $B_\mu$), $p$ refers to derivatives $\partial$
or light masses $m_{h,W,Z}$ acting appropriately
on the fields, and $f_k^{(d)}$ are the corresponding couplings
of the operator ($f_k^{(2)}\sim {\rm v}^2$, $f_k^{(4)}\sim a_i , c_\gamma...$).
Let us now consider an arbitrary diagram
with $L$ loops, $I$ internal boson propagators and $N_d$ vertices from $\mL_d$  (with
total number of vertices $V=\sum_d N_d$).
Following Weinberg's dimensional arguments~\cite{Weinberg:1979},
it is not difficult to see that
in dimensional regularization this amplitude will scale with $p$
like~\cite{photon-scat,Weinberg:1979,EW-chiral-counting}
\bear
&&\hspace*{-1cm}\mM  \,\,  \sim \,\,   \Int (d^4p)^L\, \Frac{1}{(p^2)^{I}}\, \prod_d (p^d)^{N_d}
\,\, \sim \,\,  p^{4L - 2I  + \sum_d d N_d}
\nn\\
&&\hspace*{-0.5cm}\sim \, \,\, p^{2+2L + \sum_d (d-2) N_d}\, ,
\eear
where we have used the topological identity $I=L+V-1$ in the last line.
Finally, keeping track of the constant factors with powers of  $(16\pi^2)^{-1}$ (from loops)
and ${\rm v}$ (coming with every field $\chi$ in~(\ref{eq.Ld})), and the coupling constants $f_k^{(d)}$
(from every vertex $\mL_d$), it is not difficult to complete the previous formula
into~\cite{photon-scat}
\bear
&&\hspace*{-1cm} \mM  \,\, \sim\,\,
\left(\Frac{p^2}{{\rm v}^{N_E-2}}\right)\, \left(\Frac{p^2}{16\pi^2 {\rm v}^2}\right)^L\,
\prod_d\left( \Frac{f_k^{(d)} p^{d-2}}{{\rm v}^2}\right)^{N_d}\, ,
\label{eq.counting}
\eear
with $N_E$ the number of external boson legs, which shows up in the final expression after
counting the total number of fields from all the $\mL_d$  vertices, and hence the total number
of powers of ${\rm v}^{-1}$: the diagram carries then the factor
$({\rm v}^{-1})^{ 2 I +N_E }= ({\rm v}^{-1})^{N_E-2 + 2L + \sum_d 2 N_d }$, as shown above.

The various possible contributions to the amplitude of a given process can be then sorted
in the form
\bear
\mM &=& \underbrace{ \mM_{\cO(p^2)}   }_{{\rm\bf LO} } \,\, +\,\,
\underbrace{  \mM_{\cO(p^4)}      }_{{\rm\bf NLO} }   \,\, +\,\, ...
\eear
Observing Eq.~(\ref{eq.counting}) one can see that higher orders in the chiral expansion can be reached
by either adding more loops $L$ to the diagram or vertices of ``chiral dimension'' $d\geq 4$.
Notice that adding vertices from $\mL_2$ does not modify the scaling of the diagram with $p$, as far as
the number of loops $L$ remains the same.
At LO, one needs to consider only the tree-level diagrams made out of $\mL_2$ vertices
($L=0$, $N_{2}$ arbitrary, $N_{d\geq 4}=0$);
at NLO, one needs to compute the one-loop diagrams with $\mL_2$ vertices
($L=1$, $N_{2}$ arbitrary,  $N_{d\geq 4}=0$) and tree-level
diagrams with one vertex from $\mL_4$ and any number of vertices from $\mL_2$
($L=0$, $N_{2}$ arbitrary,  $N_{4}=1$, $N_{d\geq 6}=0$); the procedure is analogous for higher
chiral orders.

\begin{figure}
\begin{center}
\psfrag{X}{$\gamma$}
\includegraphics[scale=0.35]{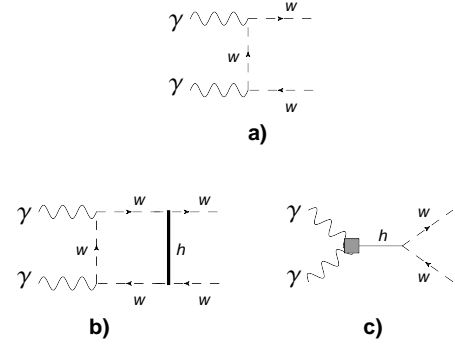}
%%%\\
%%%\vspace*{-0.75cm}
%%%\flushleft\hspace*{5cm} {\bf a)} \hspace*{8cm} {\bf b)}
%
\caption{\small{
Examples of contributing  diagrams to $\gamma\gamma\to w^+ w^-$.
{\bf a)} %%%$\cO(p^2)$
LO diagram ($L=0$ with only $\mL_2$ vertices);
{\bf b)} %%%$\cO(p^4)$
NLO loop diagram ($L=1$ with only $\mL_2$ vertices);
{\bf c)} %%%$\cO(p^4)$
NLO tree-level diagram ($L=0$ with one vertex from $\mL_4$ and any number of $\mL_2$ vertices).
The arrow in the charged $w$ lines indicates the electric charge flow. All the vertices are from $\mL_2$
but for the gray square in {\bf c)}, which comes from $\mL_4$.
}}
\label{fig.diagrams}
\end{center}
\end{figure}

In our particular computation of $\mM(\gamma\gamma\to w^a w^b)$ up to NLO, the
contributions we find are sorted out in the form~\cite{photon-scat}
\begin{equation}
\hspace*{-0.5cm}\mM\, =\,
\underbrace{\cO(e^2)}_{\rm LO,\, tree}
\, +\,  %%%\left(
\underbrace{\cO\left( e^2\Frac{p^2}{16\pi^2 {\rm v}^2}\right) }_{\rm NLO,\, 1-loop}
\, +\,
\underbrace{\cO\left( e^2 \Frac{a_i p^2}{{\rm v}^2}\right) }_{\rm NLO,\, tree}
%%% \right)
 \,  ,
 \label{eq.M}
\end{equation}
where $e\sim \cO(p/{\rm v})$ and $a_i$ stands for a general $\mL_4$ coupling.
These three types of contributions can be better understood through the detailed analysis of the
examples in Fig.~\ref{fig.diagrams}, three of the many diagrams entering in
$\gamma\gamma\to w^+ w^-$ up to NLO~\cite{photon-scat}:
\begin{itemize}

\item{\bf a)} The tree-level amplitude in Fig.~\ref{fig.diagrams}.a  with vertices from $\mL_2$
scales like~\cite{photon-scat}
$$\mM_{{\rm\bf a}}\sim \, (e\, p)\, \Frac{1}{p^{2}}\, (e\, p)  \,\sim \, e^2\, ,$$
with each $\gamma w^+w^-$
vertex scaling like $e\, p$ and the intermediate propagator like $p^{-2}$.

\item{\bf b)} The one-loop amplitude in Fig.~\ref{fig.diagrams}.b  with vertices from $\mL_2$
scales like~\cite{photon-scat}
$$\mM_{{\rm\bf b}}\sim \int \Frac{d^4p}{(2\pi)^d} \, \, (e\, p)^2 \,
\left(\frac{p^2}{{\rm v}}\right)^2\, \Frac{1}{(p^{2})^4}
\,\sim\, e^2\, \Frac{ p^2}{16\pi^2 {\rm v}^2}\, ,$$
with each $\gamma w^+w^-$ vertex scaling like $e\, p$, each $hw^+w^-$ vertex like $p^2/{\rm v}$ and
each internal propagator like $p^{-2}$.
This amplitude actually comes together with logarithms of the energy  and
ultraviolet (UV) divergences.

\item{\bf c)} The tree-level amplitude in Fig.~\ref{fig.diagrams}.c  with one vertex from $\mL_4$
and vertices from $\mL_2$ scales like~\cite{photon-scat}
$$\mM_{{\rm\bf c}}\sim \, \, \left(\Frac{ c_\gamma \, e^2 p^2}{{\rm v}}\right) \,\Frac{1}{p^2} \, \left(\Frac{p^2}{{\rm v}}\right)
\,\sim\, e^2\, \Frac{ c_\gamma p^2}{{\rm v}^2}\, ,$$
with the $\gamma\gamma h$ vertex from $\mL_4$ scaling like~\footnote{
Notice the typo in the $h\gamma\gamma$ Feynman rule in App.~A.2 in Ref.~\cite{photon-scat}, where a factor $e^2$ is missing.
}
$c_\gamma e^2 p^2/{\rm v}$,
the  $hw^+w^-$ vertex like   $p^2/{\rm v}$ and
the intermediate Higgs propagator like $p^{-2}$.
In general, the cancelation of the UV divergences in the one-loop NLO diagrams will
require the renormalization of the $\mL_4$ couplings, e.g.,  $c_\gamma^r=c_\gamma+\delta c_\gamma$,
$a_i^r= a_i+\delta a_i$.

\end{itemize}

%%%%%%%%%%%%%%%%%%%%%%%%%%%%%%%%%%%%%%%%%%%%%%%%%%%%%%%%%%%%%%%
%%%%%%%%%%%%%%%%%%%%%%%%%%%%%%%%%%%%%%%%%%%%%%%%%%%%%%%%%%%%%%%

The $\mM(\gamma(k_1,\epsilon_1)\gamma(k_2,\epsilon_2)\to w^a(p_1) w^b(p_2))$ amplitudes,
with $w^a w^b=zz,w^+w^-$, have the Lorentz decomposition~\cite{photon-scat,photon-scat-ChPT,Morales:92}
\be
\hspace*{-0.5cm}\mM   =
ie^2 (\epsilon_1^\mu \epsilon_2^\nu T_{\mu\nu}^{(1)}) A(s,t,u)+
ie^2 (\epsilon_1^\mu \epsilon_2^\nu T_{\mu\nu}^{(2)})B(s,t,u),
\ee
written in terms of the two independent Lorentz structures
$T_{\mu\nu}^{(1)}\sim\cO(p^2)$ and $T_{\mu\nu}^{(2)}\sim\cO(p^4)$
involving the external momenta,
%%%which can be found in~\cite{photon-scat}.
%%%,
\bear
&&\hspace*{-1.cm}
(\epsilon_1^\mu\epsilon_2^\nu T^{(1)}_{\mu\nu}) =
\frac{s}{2} (\epsilon_1 \epsilon_2) - (\epsilon_1 k_2)(\epsilon_2 k_1),
\label{eq.T1-def}
\\
&&\hspace*{-1.cm}
(\epsilon_1^\mu\epsilon_2^\nu T^{(2)}_{\mu\nu}) =
2 s (\epsilon_1 \Delta)(\epsilon_2 \Delta)
- (t-u)^2 (\epsilon_1\epsilon_2)
\nn\\
&&\hspace*{1cm}
- 2 (t-u) [(\epsilon_1\Delta)(\epsilon_2 k_1) - (\epsilon_1 k_2) (\epsilon_2 \Delta)]\, ,
\nn
%%%\label{eq.T2-def}
\eear
with $\Delta^\mu\equiv p_1^\mu -p_2^\mu$.
%%%
The Mandelstam variables are defined as
$s=(p_1+p_2)^2$, $t=(k_1-p_1)^2$ and $u=(k_1-p_2)^2$ and the $\epsilon_i$'s
are the  polarization vectors of the initial photons.
At LO and NLO we find for the neutral channel~\cite{photon-scat},
\bear
&&\hspace*{-1.5cm}
A(\gamma\gamma\to zz)_{\rm LO}
= B(\gamma\gamma\to zz)_{\rm  LO}
= B(\gamma\gamma\to zz)_{\rm  NLO}
= 0 \, ,
\label{eq.photon-scat}
\nn  \\
&&\hspace*{-1.5cm}
A(\gamma \gamma \to zz)_{\rm NLO}=
\Frac{2 a c_\gamma^r}{{\rm v}^2} + \Frac{ (a^2-1)}{4\pi^2{\rm v}^2}\, ,
\label{eq.A-zz}
\label{eq.results-zz}
\eear
and for the charged one~\cite{photon-scat}
\bear
&&\hspace*{-1.5cm}
A(\gamma\gamma\to w^+ w^-)_{\rm LO}    %%%^{\rm Born}
= 2 s B(\gamma\gamma\to w^+ w^-)_{\rm  LO}    %%%^{\rm Born}
= -\Frac{1}{t} - \Frac{1}{u},
\nn\\
&&\hspace*{-1.5cm}
A(\gamma \gamma \to w^+w^-)_{\rm NLO}
\hspace*{-0.1cm}
=
\hspace*{-0.1cm}
\Frac{2 a c_\gamma^r}{{\rm v}^2}
+ \Frac{8(a_1^r-a_2^r+a_3^r)}{{\rm v}^2}
 + \Frac{ (a^2-1)}{8\pi^2{\rm v}^2} ,
\label{eq.A-zz}
%%%\\
\nn \\
&&\hspace*{-1.5cm}
B(\gamma \gamma \to w^+w^-)_{\rm NLO} =0 \, .
 \label{eq.results-ww}
\eear
The term with $c_\gamma^r$  comes from the Higgs tree-level
exchange in the $s$--channel, the term proportional to $(a^2-1)$
comes from the one-loop diagrams with $\mL_2$ vertices,
%%%(Fig.~\ref{ggtozz}).
and the Higgsless operators in~(\ref{eq.L4}) yield the tree-level contribution
to $\gamma\gamma\to w^+ w^-$ proportional to $(a_1-a_2+a_3)$. 
Independent diagrams are  in general UV divergent and have complicated logarithmic and Lorentz
structure.~\footnote{
For instance, the diagram shown in Fig.~\ref{fig.diagrams}.c corresponds to the diagram~14
in App.~B.2 in Ref.~\cite{photon-scat}, given by the complicate structure
\begin{eqnarray}
%%%\mM^{\rm 3} &=& \frac{i a^2 e^2 }{288 \pi ^2 {\rm v}^2}
%%%\bigg(3 B_0{(t,0,0)} (2 \left(\epsilon _1\epsilon _2\right) t
%%%-5 (\left(\epsilon _1{$\Delta $)}\right.+\left(\epsilon _1k_2\right))
%%%(\left(\epsilon _2{$\Delta $)}\right.-\left(\epsilon _2k_1\right)))
%%%\nn\\
%%%&&\qquad\quad +(-\left(\epsilon _1{$\Delta $)}\right.-\left(\epsilon _1k_2\right)) (\left(\epsilon _2{$\Delta $)}\right.-\left(\epsilon _2k_1\right))+4 \left(\epsilon _1\epsilon _2\right) t\bigg)
%%%\, ,\\
&&\hspace*{-1.4cm}
\mM^{\rm 14} = - \frac{i a^2 e^2 }{288 \pi ^2 s {\rm v}^2}
\bigg(6 (-s) (B_0 (s,0,0)  (\left(\epsilon _1  \Delta  ) \right.
\left(\epsilon _2k_1\right)-\left(\epsilon _1k_2\right)
 \left(\epsilon _2  \Delta  ) \right.
\nn\\
&&
\hspace*{-1.3cm}
 +2 \left(\epsilon _1\epsilon _2\right) t
 +\left(\epsilon _1\epsilon _2\right) u)+B_0{(t,0,0)}
  ((\left(\epsilon _1 \Delta  ) \right.+\left(\epsilon _1k_2\right))
  (\left(\epsilon _2 \Delta  ) \right.-\left(\epsilon _2k_1\right))
\nn\\
&&
\hspace*{-1.5cm}
  -\left(\epsilon _1\epsilon _2\right) t))
 +\left(\epsilon _1 \Delta  ) \right.
  (\left(\epsilon _2 \Delta  ) \right.+3 \left(\epsilon _2k_1\right)) (-s)
  +\left(\epsilon _1k_2\right) (\left(\epsilon _2k_1\right) (23 t+11 u)
\nn\\
&&\qquad\quad
\hspace*{-1.5cm}
  -3 \left(\epsilon _2 \Delta  ) \right. (-s))
  +2 \left(\epsilon _1\epsilon _2\right) (5 t+2 u) (-s)\bigg)
\, ,
\nn
\end{eqnarray}
}
However, in dimensional regularization, when all the different contributions
(10 and 39 loop diagrams for the neutral and charged channels, respectively)
are put together
the final one-loop amplitude turns out to be UV finite and free of logs
in the limits considered in our analysis~\cite{photon-scat}, both in $\gamma\gamma\to zz$
and $\gamma\gamma\to w^+ w^-$.  Therefore the combinations of NLO couplings $c_\gamma$ and
$(a_1-a_2+a_3)$ which enter here  do not need to be renormalized:
$ a_1^r-a_2^r+a_3^r=a_1 -a_2 +a_3 $  (like in the Higgsless case~\cite{photon-scat-ChPT,Morales:92})
and  $c^r_\gamma=c_\gamma$ are renormalization group invariant~\cite{photon-scat}.
All the UV divergences and renormalizations occur at $\cO(p^4)$ and the $\mL_2$ couplings
(like $a$, for instance) do not get renormalized within the approximations considered
in this work~\cite{photon-scat}.

\begin{table}[!t]
\begin{center}
\begin{tabular}{l|cc}
%\hline\hline
    & \multicolumn{2}{c}{\bf Relevant combinations} \\
 {\bf Observables}  & \multicolumn{2}{c}{\bf  of parameters} \\
\null  & \ \ from $\mL_2$\ \  & from $\mL_4$ \\
\hline
$\mM(\gamma\gamma\to zz)$     & $a$ & $c_\gamma^{r}$ \\
$\mM(\gamma\gamma\to w^+w^-)$ & $a$ & $(a_1^r- a_2^r+a_3^r),\, c_\gamma^{r}$ \\
$\Gamma(h\to\gamma\gamma)$    & $a$ & $c_\gamma^{r}$ \\
$S$--parameter                & $a$ &
    $a_1^r$
\\
$\mF_{\gamma^*ww}$            & $a$ &
      $(a_2^r-a_3^r)$
\\
$\mF_{\gamma^*\gamma h}$      & --  & $c_\gamma^{r}$\\
\hline  %\hline
\end{tabular}
\caption{{\small
Set of observables studied in Ref.~\cite{photon-scat} and their corresponding  relevant
combinations of chiral parameters. 
}}
\vspace*{-0.5cm}
\label{tab.relevant-couplings}
\end{center}
\end{table}

\begin{table}[!t]
\begin{center}
\begin{tabular}{ c|c|c }
%%%\hline \hline
\rule{0pt}{3ex}
 & {\bf ECLh  } &  {\bf ECL}
 \\
 &&   (Higgsless)
\\[5pt] \hline
\rule{0pt}{3ex}
$  \Gamma_{a_1-a_2+a_3} $ & $ 0  $ &  0                %%%& 0
\\[5pt] \hline
\rule{0pt}{3ex}
$  \Gamma_{c_\gamma} $ & $ 0  $ &  -                %%%& 0
\\[5pt] \hline
\rule{0pt}{3ex}
$  \Gamma_{a_1} $ & $ -\frac{1}{6}(1-a^2)  $&$ -\frac{1}{6}  $                %%%& 0
\\[5pt] \hline
\rule{0pt}{3ex}
$  \Gamma_{a_2-a_3}  $ & $ -\frac{1}{6}(1-a^2)  $& $ -\, \frac{1}{6}  $                  %%%& 0
\\[5pt] \hline
\rule{0pt}{3ex}
$  \Gamma_{a_4} $ & $ \frac{1}{6}(1-a^2)^2   $ & $ \frac{1}{6} $                 %%%& 0
\\[5pt] \hline
\rule{0pt}{3ex}
$  \Gamma_{a_5} $ & $ \frac{1}{8}(b-a^2)^2
+\frac{1}{12}(1-a^2)^2 $ &
$  \frac{1}{12} $                 %%%& 0
\\[5pt] \hline
%%%\hline
\end{tabular}
\caption{\small
Running of the relevant ECLh parameters and their combinations appearing
in the six selected observables~\cite{photon-scat}. 
The third column provides the corresponding running
for the Higgsless EW Chiral Lagrangian~\cite{Morales:94}.
The table has been completed with the running of $a_4$ and $a_5$ from $WW$--scattering
analyses~\cite{WW-theory}.
}
\vspace*{-0.5cm}
\label{tab.running}
\end{center}
\end{table}

Our $\gamma\gamma$--scattering amplitudes depend on three combinations of parameters ($a$, $c_\gamma^r$
and $a_1^r-a_2^r+a_3^r$). This tells us that in order to extract each coupling separately one needs to study
other observables. However, other related photon processes are ruled by
the same parameters. In Ref.~\cite{photon-scat} we provide a list of four additional observables, computed
with the ECLh under the same assumptions of this work and depending on different combinations of
$a$, $c_\gamma^r$, $a_1^r$ and $(a_2^r-a_3^r)$: the $h\to\gamma\gamma$ partial width,
the oblique $S$--parameter
and the electromagnetic form-factors for $\gamma^*\to w^+ w^-$ and $\gamma^*\to h \gamma$.
In table~\ref{tab.relevant-couplings} one can see the combinations of couplings that rule each quantity.
This gives six observables and four relevant combinations. Thus,
the ECLh allows us to extract the couplings from four observables and
make a definite prediction for the other two. Notice that a global fit with the non-linear EFT
must incorporate both NLO loops and NLO tree-level contributions
(both are of the same order in the chiral counting),
otherwise one may eventually run into inconsistent determinations.

These six observables provide in addition a consistent set of renormalization conditions
($a_1$ and $a_2-a_3$ do need to be renormalized). The corresponding running for the $\cO(p^4)$
couplings $C^r=c_\gamma^r,a_i^r$ are summarized in Table~\ref{tab.running}, where the constants $\Gamma_C$ therein
are given by
\bear
\Frac{d C^r}{d\ln\mu} &=& \,-\,\Frac{\Gamma_C}{16\pi^2}\,.
\eear
For the sake of completeness, we have also included in the last two lines of Table~\ref{tab.running}
the running of $a_4^r$ and $a_5^r$ determined in $WW$--scattering analyses~\cite{WW-theory}.

A remarkable feature of the one-loop photon-photon amplitudes
is that individual diagrams carry the usual chiral  suppression
$\cO\left(p^2/(16\pi^2 {\rm v}^2)\right)$  with respect to the LO.
However, the full one-loop amplitude shows a stronger suppression
$\cO\left((1-a^2)p^2/(16\pi^2 {\rm v}^2)\right)$, where experimentally $a$ is found to be close to $1$
within $\cO(10\%)$ uncertainties~\cite{LHC-exp}.

\begin{figure}
\begin{center}
\includegraphics[scale=0.5]{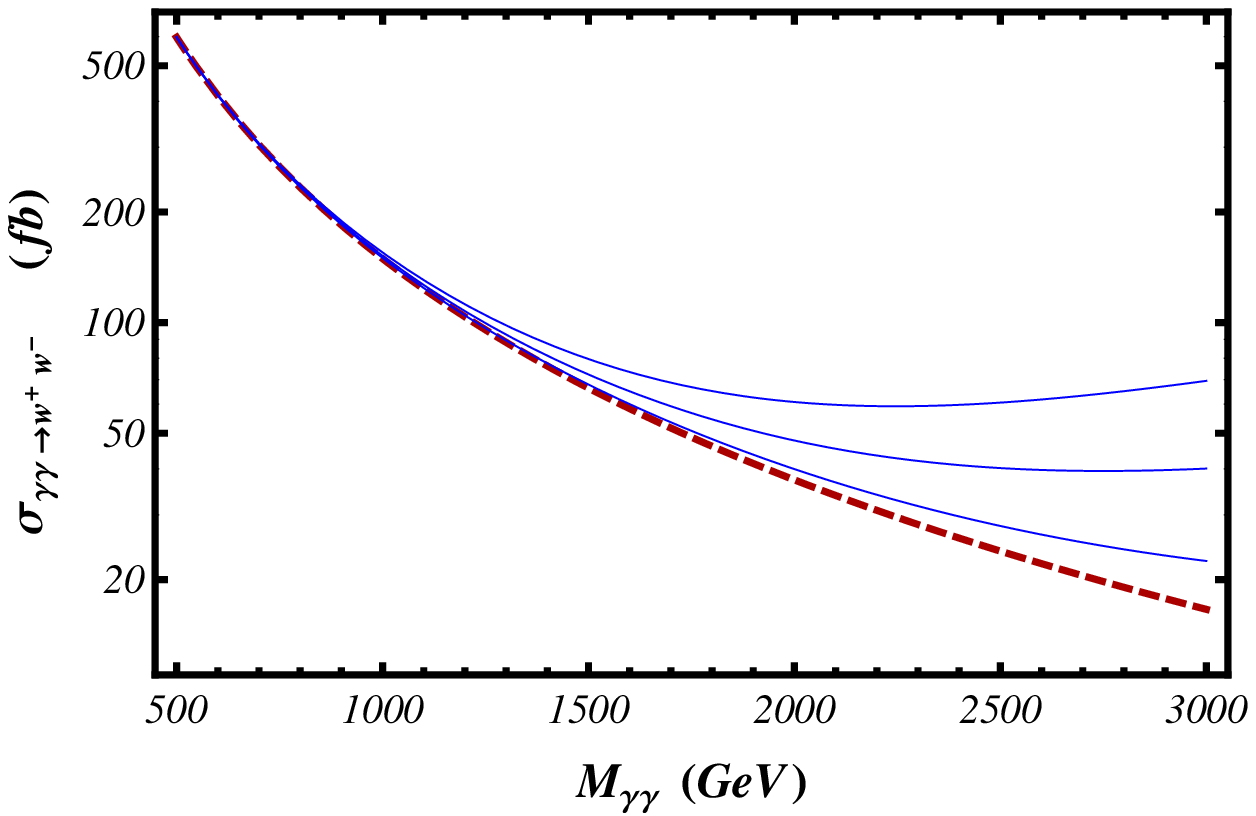}  \\
\includegraphics[scale=0.5]{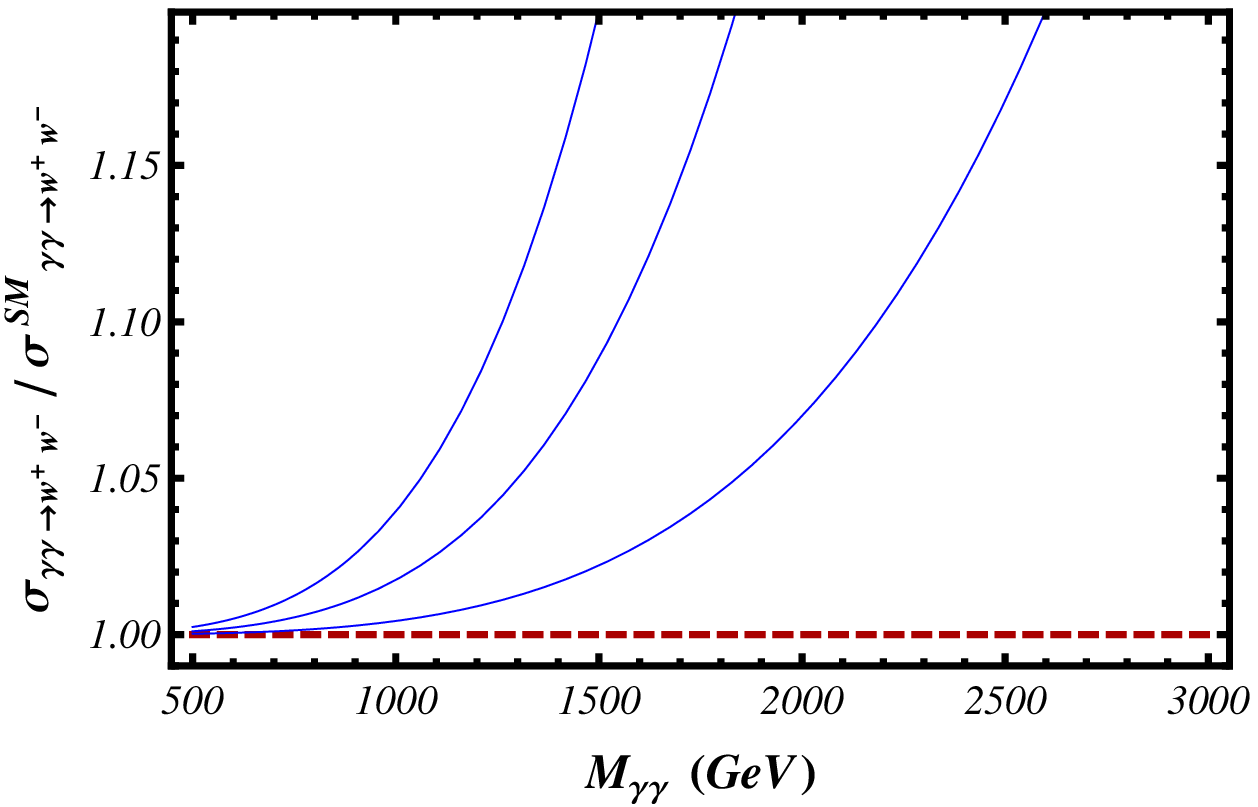}
%%%\\
%%%\vspace*{-0.75cm}
%%%\flushleft\hspace*{5cm} {\bf a)} \hspace*{8cm} {\bf b)}
%
\caption{\small{.
Cross section (top) for $\gamma\gamma\to W_L^+ W_L^-$ for unpolarized photons.
The ratio of the ECLh and SM cross sections is provided in the lower plot.
The red-dashed line correspond to the SM prediction
and the solid blue ones our ECLh predictions for $a=c_\gamma=0$
and $(a_1-a_2+a_3)= 2\times 10^{-3},\,4\times 10^{-3},\,6\times 10^{-3}$,
respectively from bottom to top in each plot.
}}
\end{center}
\label{fig.sigma}
\end{figure}

We would like to finish this section with the preliminary phenomenological analysis for
$\gamma\gamma\to W_L^+ W_L^-$ shown in Fig.~\ref{fig.sigma}.
The fact that the Equivalence Theorem works with an error lower that 2\%
in the SM for $M_{\gamma\gamma}=\sqrt{s}>0.5$~TeV reassures us about the validity of our analysis.
The SM cross section behaves at high energies like $1/s$ for $\gamma\gamma\to W_L^+ W_L^-$.
On the other hand, the $\cO(p^4)$ NLO terms in the amplitude~(\ref{eq.results-ww})
add a contribution to the cross section that grows with $s$ and turns more and more important at higher
and higher energies. We observe the impact of possible new physics by varying the couplings
within typical ranges for the chiral couplings~\cite{chpt,Morales:92}:
$a_1^r-a_2^r+a_3^r= 2\times 10^{-3},\, 4\times 10^{-3},\, 6\times 10^{-3}$
(respectively from bottom to top in Fig.~\ref{fig.sigma}),
and the other couplings set to their SM values, $a=1$ and $c_\gamma^r=0$.
The deviation from the SM is negligible at very low energies. Nonetheless,
it grows with $M_{\gamma\gamma}$
and for $a_1-a_2+a_3=2\times 10^{-3}$  ($4\times 10^{-3}$; $6\times 10^{-3}$)
the cross section exceeds the SM one by 20\% for $M_{\gamma\gamma}>2.6$~TeV  (1.8~TeV; 1.5~TeV).
The signal keeps turning more and more intense beyond these values of $M_{\gamma\gamma}$.
A more detailed study will be provided in a forthcoming work.
In order to study this subprocess in colliders (LHC or future $e^+ e^-$ accelerators)
we will have to convolute this $\gamma\gamma$ cross sections
with the corresponding photon luminosity functions.
Although preliminary studies show that one can get a measurable amount of events for integrated
luminosities of the order of 1~ab$^{-1}$, the key-point will be the discrimination and separation
of SM background through convenient cuts~\cite{photon-scat-CMS,photon-scat-SM,Han:2005}
and the minimization of theoretical uncertainties. 
For instance, the non-zero $h$, $W$ and $Z$ masses produce corrections suppressed by
$m_{h,W,Z}/M_{\gamma\gamma}$,
which may turn important if one studies this reaction below the TeV. This also means going beyond the
Equivalence Theorem and computing the full one-loop $\gamma\gamma\to V_L V_L$ amplitude.
It can be also interesting to analyze within this framework the reverted subprocess
$VV\to \gamma\gamma$ via vector boson fusion at LHC.

%%%%%%%%%%%%%%%%%%%%%%%%%%%%%%%%%%%%%%%%%%%%%%%%%%%%%%%%%%%%%%%
%%%%%%%%%%%%%%%%%%%%%%%%%%%%%%%%%%%%%%%%%%%%%%%%%%%%%%%%%%%%%%%

\section{$\gamma\gamma$--scattering in MCHM}

In this section we show an explicit example of how our EFT description describes the small momentum regime
of any underlying theory with the same symmetries and low-energy particle content.
%%%  (one singlet $h$, three EW WBGBs, etc.)

In the context of the so called $SO(5)/SO(4)$ MCHM~\cite{MCHM}
it is assumed that some global symmetry breaking takes place at some scale
$4 \pi f>  4\pi {\rm v}$ so that the group $G=SO(5)$ is spontaneously broken to the subgroup $H=SO(4)$.
The corresponding NGBs
live in the coset $G/H=S^4$. These four NGBs $\omega^\alpha$
are then identified with the Higgs-like boson $h$
and the three WBGBs needed for giving masses to the $W^\pm$ and $Z$
($w^\pm =(\omega^1\mp i \omega^2)/\sqrt{2}$, $z=\omega^3$, $h=\omega^4$).
%%%
%%%\bear
%%%\Phi_0' = f \left(\begin{array}{c}0 \\ 0\\ 0\\ 0 \\ 1\end{array}\right)\, ,
%%%\qquad \qquad
%%%\Phi_0 =  f \left(\begin{array}{c}0 \\ 0\\ 0\\ s \\ c \end{array}\right)\, ,
%%%\eear
%%%

The low-energy dynamics of the MCHM NGBs and the EW gauge bosons can be described
through  the gauged non-linear sigma model (NL$\sigma$M)~\cite{photon-scat,MCHM}
(only operators with photons and NGBs are shown here),
\bear
  \mL^{\rm MCHM}_2
&=& \Frac{1}{2} D^\mu \Phi^{\,\, \dagger} \, D_\mu\Phi\, \mid_{S^4}
\label{eq.Phi-vector}
\nn\\
&&
%%%
%%%= \Frac{1}{2}\, g_{\alpha\beta}(\omega)\,  D^\mu \omega^\alpha\, D_\mu \omega^\beta
%%%\nn \\  &&
%%%
\hspace*{-1.1cm}
= \frac{1}{2}\, g_{\alpha\beta}(\omega)\,  \partial^\mu \omega^\alpha\, \partial_\mu \omega^\beta
\nn\\
&&
\hspace*{-0.9cm}
+ \, i\, e A_\mu(\omega^-\partial^\mu \omega^+-\omega^+\partial^\mu\omega^-)+   e^2 A^2 \omega^+\omega^-.
\nn
\eear
with the $G$--fundamental representation vector $\Phi$ parametrizing the NGBs in the way
\bear
&&
\hspace*{-1.5cm}
\Phi = \left(\begin{array}{c}\omega^1 \\ \omega^2\\ \omega^3\\ c \omega^4+s \chi
\\  -s\omega^4+ c\chi \end{array}\right), \,\,
\mbox{with } \chi= \left(f^2 -\sum_{\alpha=1}^{4} (\omega^\alpha)^2\right)^{1/2} ,
\eear
with  $s= \sin \theta$,   $c=\cos \theta $ and  $\theta $ being the vacuum misalignment angle, with
$ \sin\theta={\rm v}/f$~\cite{MCHM}.
The $S^4$ metric is  given by
\bear
 g_{\alpha\beta}  & =  &
 \delta_{\alpha\beta}+\frac{\omega^\alpha\omega^\beta}{f^2-\sum_\alpha (\omega^\alpha)^2} \, .
\eear

%%%%%%%%%%%%%%%%%%%%%%%%%%%%%%%%%%%%%%%%%%%%%%%%%%%%%%%%%%%%%%%
\begin{figure}[!t]
%\begin{table}[ht]
\begin{center}
\includegraphics[scale=0.75]{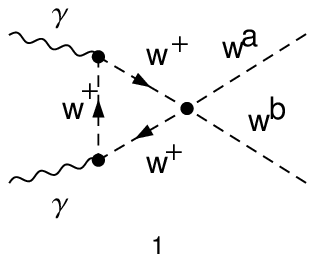}
\includegraphics[scale=0.75]{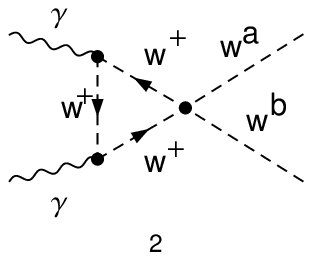}
\includegraphics[scale=0.75]{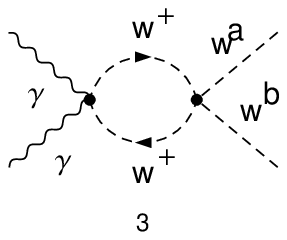}
%%%
%%%\psfig{file=MCHM-diag1.eps,width=2.5cm,clip=} & \psfig{file=MCHM-diag2.eps,width=2.5cm,clip=}
%%%& \psfig{file=MCHM-diag3.eps,width=2.5cm,clip=}
%%%
\caption{{\small  MCHM one-loop diagrams for $\gamma\gamma\to w^a w^b$ at NLO. }}
\label{fig.MCHM}
\end{center}
%\end{table}
\end{figure}

The $\gamma\gamma$--scattering was considered in the framework of general
$SO(N+1)/SO(N)$ gauged NL$\sigma$M~\cite{DobadoMorales} for low-energy QCD  and
the one-loop computation only involves the  bubble and triangle diagrams
(Fig.~\ref{fig.MCHM}). The one-loop result at NLO is simply
\bear
&&\hspace*{-1.5cm}
A(\gamma\gamma\to zz)^{\rm NLO-loop} = A(\gamma\gamma\to hh)^{\rm NLO-loop}
= \,-\, \Frac{1}{4\pi^2 f^2}
%%%\,=  \,-\, \Frac{(1-a^2)}{4\pi^2 {\rm v}^2}
\, ,
\nn\\
&&\hspace*{-1.5cm}
A(\gamma\gamma\to w^+ w^-)^{\rm NLO-loop} = \,-\, \Frac{1}{8\pi^2 f^2}
%%%\, = \,-\,\Frac{(1-a^2)}{8\pi^2 {\rm v}^2}
\, ,
\nn\\
\label{eq.results2}
\eear
and $B^{\rm NLO-loop}=0$ in all cases. We find this in agreement with
our ECLh result in Eqs.~(\ref{eq.results-zz}) and~(\ref{eq.results-ww})
by means of the relation $(1-a^2)={\rm v}^2/f^2$ between $f$, ${\rm v}$ and the $hW^+W^-$ coupling $a$
in the  $SO(5)/SO(4)$ MCHM~\cite{MCHM}.

We want to remark that the $\mL_2^{\rm MCHM}$ is often written in exponential coordinates~\cite{MCHM}
rather than the $S_4$ parametrization used in this computation~\cite{photon-scat}, leading to a low-energy
Lagrangian with exactly the same structure as $\mL_2$ in Eq.~(\ref{eq.ECLh}) but with
precise predictions for the ECLh couplings. One can then use this Lagrangian in terms of exponential
coordinates and compute the $\gamma\gamma$--scattering in the way done in this work
(in that same parametrization), with all its complication and tricky diagrammatic cancelations.
The final outcome agrees with~(\ref{eq.results2}), as expected. The lesson one draws is that, even though
all the coset parametrizations yield the same outcome for a given on-shell amplitude,
computations can be simpler for some choices of the NGB coordinates
(we already saw this in our ECLh calculation
in the previous section, where some vertices are absent in spherical coordinates
and one has fewer diagrams to compute~\cite{photon-scat}).
Likewise, in the exponential parametrization individual
loop diagrams are suppressed with respect to the LO by $\cO\left( p^2/(16\pi^2 {\rm v}^2)\right)$
%%%
%%%(since experimentally $a\sim 1$)
%%%
and only after summing up all of them one finds that the full one-loop amplitude is actually suppressed
by $\cO\left( p^2/(16\pi^2 f^2)\right)$.
On the other hand, in the $S_4$ coordinates each single diagram shown in Fig.~\ref{fig.MCHM}
already carries the final suppression
$p^2/(16\pi^2 f^2)$ with respect to the LO.

\section*{Acknowledgements}

We would like to thank the organizers for the nice scientific environment
during the conference.
This work has been partly supported by
the European Union FP7 ITN INVISIBLES (Marie Curie Actions, PITN- GA-2011- 289442),
by the CICYT through the projects FPA2012-31880, FPA2010-17747,    CSD2007-00042,
FPA2011-27853-C02-01
and FPA2013-44773-P,   %%% NEW FUND w/ J.Moreno
by the CM (Comunidad Autonoma de Madrid) through the project HEPHACOS S2009/ESP-1473,
by the Spanish Consolider-Ingenio 2010 Programme CPAN (CSD2007-00042)
and by the Spanish MINECO's grant BES-2012-056054
and "Centro de Excelencia Severo Ochoa" Programme
under grant SEV-2012-0249.



%% References


%%%%%
%%%%%\bibliographystyle{elsarticle-num}
%%%%%\bibliography{<your-bib-database>}

\begin{thebibliography}{00}



\bibitem{LHC-exp}
%%%\bibitem{ATLAS-exp}
ATLAS Collaboration, Report No. ATLAS-CONF-2014-009;
%%%
%%%\bibitem{CMS-exp}
CMS Collaboration, Report No. CMS-PAS-HIG-14-009.


\bibitem{photon-scat}
%%%One-loop γγ→ W+L W−L and γγ→ ZL ZL from the Electroweak Chiral Lagrangian with a light Higgs-like scalar
    R.L. Delgado, A. Dobado, M.J. Herrero, J.J. Sanz-Cillero,
    JHEP 1407 (2014) 149
    [arXiv:1404.2866 [hep-ph]].



\bibitem{Weinberg:1979}
%%%    Phenomenological Lagrangians,
    S.  Weinberg,
    Physica A{\bf 96} (1979) 327.


\bibitem{chpt}
%%\bibitem{ChPTp4}
  J.~Gasser and H.~Leutwyler,
%%%  {\it Chiral perturbation theory to one loop},
  Annals Phys. {\bf 158} (1984) 142;
%%%    {\it Chiral perturbation theory: expansions in the mass of the strange quark},
    Nucl.\ Phys.\ B {\bf 250} (1985) 465;
%%%  {\it Low-energy expansion of meson form factors},
  Nucl.\ Phys.\  B {\bf 250} (1985) 517.
  %%CITATION = NUPHA,B250,517;%%




\bibitem{Machado:2014}
%%%On the renormalization of the electroweak chiral Lagrangian with a Higgs
    M.B. Gavela, K. Kanshin, P.A.N. Machado and S. Saa,
    [arXiv:1409.1571 [hep-ph]].


\bibitem{Jikia:1993}
%%%Z boson pair production via photon fusion at a high-energy photon linear collider
    G. Jikia,
    Nucl.Phys. B{\bf 405} (1993) 24.   %%%-54



\bibitem{equivalence-theorem}
    J. M. Cornwall, D. N. Levin and  G. Tiktopoulos,
%%%    {\it Derivation of Gauge Invariance from High-Energy Unitarity Bounds on the S Matrix},
    Phys.Rev. D {\bf 10} (1974) 1145, Erratum-ibid.  D {\bf 11} (1975) 972;
%
    C.E. Vayonakis,
%%%    {\it Born Helicity Amplitudes and Cross-Sections in Nonabelian Gauge Theories},
    Lett.Nuovo Cim. {\bf 17} (1976) 383;
%
    B.W. Lee, C. Quigg and  H.B. Thacker,
%%%    {\it Weak Interactions at Very High-Energies: The Role of the Higgs Boson Mass},
    Phys.Rev. D {\bf 16} (1977) 1519;
%
    G.J. Gounaris,  R. Kogerler and  H. Neufeld,
%%%    {\it     Relationship Between Longitudinally Polarized Vector Bosons and their Unphysical Scalar Partners},
    Phys.Rev. D {\bf 34} (1986) 3257;
%

\bibitem{MCHM}
K. Agashe, R. Contino and A. Pomarol, Nucl. Phys. B {\bf 719}, 165 (2005)
[arXiv:hep-ph/0412089];
  R. Contino, L. Da Rold and A. Pomarol, Phys. Rev. D {\bf   75}, 055014
(2007) [arXiv:hep-ph/0612048];
    R. Contino,  D.  Marzocca,  D. Pappadopulo and R. Rattazzi,
JHEP {\bf 1110} (2011) 081 [arXiv:1109.1570 [hep-ph]];
D. Barducci et al. JHEP {\bf 1309}, 047 (2013) [arXiv:1302.2371
[hep-ph]].



%\cite{Appelquist:1980vg}
\bibitem{Appelquist:1980vg}
  T.~Appelquist and C.~W.~Bernard,
  %``Strongly Interacting Higgs Bosons,''
  Phys.\ Rev.\ D {\bf 22} (1980) 200.
  %%CITATION = PHRVA,D22,200;%%
%\cite{Appelquist:1980ix}
%%%\bibitem{Appelquist:1980ix}
%  T.~Appelquist,
%  %``Broken Gauge Theories And Effective Lagrangians,''
%  In {\it St. Andrews 1980, Proceedings, Gauge Theories and Experiments At High Energies}, p.~385.


\bibitem{Longhitano:1980iz}
  A.~C.~Longhitano,
  %``Heavy Higgs Bosons in the Weinberg-Salam Model,''
  Phys.\ Rev.\ D {\bf 22} (1980) 1166;
  %%CITATION = PHRVA,D22,1166;%%
%
%\cite{Longhitano:1980tm}
%\bibitem{Longhitano:1980tm}
%  A.~C.~Longhitano,
  %``Low-Energy Impact of a Heavy Higgs Boson Sector,''
  Nucl.\ Phys.\ B {\bf 188} (1981) 118.
  %%CITATION = NUPHA,B188,118;%%




\bibitem{ECLh-Op4}
%%%  \bibitem{Alonso:2012}
%%%    The Effective Chiral Lagrangian for a Light Dynamical 'Higgs'
    R. Alonso,  M.B. Gavela, L. Merlo, S. Rigolin and  J. Yepes,
    Phys.Lett. B{\bf 722} (2013) 330  %%%-335
    [arXiv:1212.3305 [hep-ph]];
%
%%% \bibitem{Brivio:2013}
    I. Brivio {\it et al.},
%%%(Madrid, Autonoma U. & Madrid, IFT), T. Corbett (YITP, Stony Brook & Stony Brook U.), O.J.P. Éboli (Sao Paulo U.), M.B. Gavela (Madrid, Autonoma U. & Madrid, IFT & Madrid, IFT & Madrid, Autonoma U.), J. Gonzalez-Fraile (Barcelona U., ECM & ICC, Barcelona U. & Barcelona U., ECM), M.C. Gonzalez-Garcia (ICREA, Barcelona & Barcelona U., ECM & ICC, Barcelona U. & YITP, Stony Brook & Stony Brook U.), L. Merlo (Madrid, Autonoma U. & Madrid, IFT & Madrid, IFT & Madrid, Autonoma U.), S. Rigolin (Padua U. & INFN, Padua & Padua U.). Nov 7, 2013. 63 pp.
%%%  {\it Disentangling a dynamical Higgs},
    JHEP {\bf 1403} (2014) 024
    [arXiv:1311.1823 [hep-ph]].


\bibitem{Pich:2013}
    A. Pich, I. Rosell and J.J. Sanz-Cillero,
%%%    {\it Viability of strongly-coupled scenarios with a light Higgs-like boson},
    Phys.Rev.Lett. {\bf 110}  (2013) 181801
    [arXiv:1212.6769];
%
%%%A. Pich (Valencia U., IFIC), I. Rosell (Valencia U., IFIC & Cardenal Herrera U.), J.J. Sanz-Ciller (Madrid, IFT). Oct 11, 2013. 35 pp.
%%%    {\it Oblique S and T Constraints on Electroweak Strongly-Coupled Models with a Light Higgs},
    JHEP {\bf 1401} (2014) 157
    [arXiv:1310.3121 [hep-ph]]


\bibitem{EW-chiral-counting}
%%%Chiral Quarks and the Nonrelativistic Quark Model
    A. Manohar and H. Georgi,
    Nucl.Phys. B{\bf 234} (1984) 189;
%
%%%Lepton-number violation and right-handed neutrinos in Higgs-less effective theories
    J. Hirn and J. Stern,
    Phys.Rev. D{\bf 73} (2006) 056001
    [arXiv:hep-ph/0504277];
%
%%%Effective Theory of a Dynamically Broken Electroweak Standard Model at NLO
    G. Buchalla and  O. Cata,
    JHEP {\bf 1207} (2012) 101
    [arXiv:1203.6510 [hep-ph]].



\bibitem{chpt+photons}
%%%Virtual photons in chiral perturbation theory
    R. Urech,
    Nucl.Phys. B{\bf 433} (1995) 234    %%%-254
    [arXiv:hep-ph/9405341]



\bibitem{photon-scat-ChPT}
%\bibitem{Bijnens:1988}
%%% Two Pion Production in Photon-Photon Collisions
    J.~Bijnens and F.~Cornet,
    Nucl. Phys. B \textbf{296} (1988) 557;
%
%\bibitem{Donoghue}
    J. F. Donoghue, B. R. Holstein and  Y.C. Lin,
    Phys.Rev. D {\bf 37} (1988) 2423;
%
%\bibitem{Dawson:91}
%%%    $\gamma \gamma \to  \pi^0 \pi^0$ and $K_L \to  \pi^0 \gamma \gamma$ in the chiral quark model,
    J. Bijnens,  S. Dawson and G. Valencia,
    Phys. Rev. D {\bf 44} (1991) 3555.   %%%-3561


\bibitem{Morales:92}
%%%Study of gamma gamma ---> W+(L) W-(L) and gamma gamma ---> Z(L) Z(L) reactions with chiral Lagrangians
    M. J. Herrero and E. Ruiz-Morales,
    Phys.Lett. B{\bf 296} (1992) 397   %%%-407
    [arXiv:hep-ph/9208220].





%%
%%  bibliographic items can be constructed using the LaTeX format in SPIRES:
%%    see    http://www.slac.stanford.edu/spires/hep/latex.html
%%  SPIRES will also supply the CITATION line information; please include it.
%%

\bibitem{WW-theory}
%%%\bibitem{Espriu:2013B}
    D. Espriu and B. Yencho,
    Phys. Rev D {\bf 87} (2013) 055017
    [arXiv:1212.4158 [hep-ph]];
%
    D. Espriu, F. Mescia and  B. Yencho,
    Phys. Rev D {\bf 88} (2013) 055002
    [arXiv:1307.2400 [hep-ph]];
%
   D. Espriu and B. Mescia,
   Phys.Rev. D{\bf 90} (2014) 015035
    [arXiv:1403.7386 [hep-ph]];
%%%
%%%\bibitem{Dobado:2013}
%%%     ``Light "Higgs", yet strong interactions'',
     R. L. Delgado, A. Dobado, F. J. Llanes-Estrada,
     J.Phys. G{\bf 41} (2014) 025002
     [arXiv:1308.1629 [hep-ph]];
%	
%%%    One-loop $W_L W_L$ and $Z_L Z_L$ scattering from the Electroweak Chiral Lagrangian with a light Higgs-like scalar,
%%%    R. L. Delgado, A. Dobado and  F. J. Llanes-Estrada (Univ. Complutense de Madrid). Nov 23, 2013. 21 pp.
    JHEP {\bf 1402} (2014) 121
    [arXiv:1311.5993 [hep-ph]].


%%%\bibitem{WW-exp}
%%%Evidence for Electroweak Production of W±W±jj in pp Collisions at s√=8 TeV with the ATLAS Detector
%%%    G. Aad {\it et al.} (ATLAS Collaboration)
%%%    [arXiv:1405.6241 [hep-ex]].



\bibitem{Morales:94}
%%%The Electroweak chiral Lagrangian for the Standard Model with a heavy Higgs
    Maria J. Herrero and  Ester Ruiz Morales,
    Nucl.Phys. B{\bf 418} (1994) 431-455
    [arXiv:hep-ph/9308276].



\bibitem{photon-scat-CMS}
%%%Study of exclusive two-photon production of W+W− in pp collisions at s√=7 TeV and constraints on anomalous quartic gauge couplings
    S. Chatrchyan {\it et al.}  (CMS Collaboration),
    JHEP {\bf 1307} (2013) 116
    [arXiv:1305.5596 [hep-ex]].


\bibitem{photon-scat-SM} 	
%%%W+W− pair production in proton-proton collisions: small missing terms
    M. Luszczak, A. Szczurek and C. Royon,
    [arXiv:1409.1803 [hep-ph]].


\bibitem{Han:2005}
%%%Anomalous gauge couplings of the Higgs boson at high energy photon colliders
    T. Han, Y.-P. Kuang and B. Zhang,
    Phys.Rev. D73 (2006) 055010
    [arXiv:hep-ph/0512193].




\bibitem{DobadoMorales}
%%%    A Note on the $\gamma \gamma \to  \pi^0 \pi^0$ reaction in the $1/N$ expansion of $\chi$PT,
    A. Dobado and J. Morales,
    Phys.Lett. B {\bf 365} (1996) 264  %%%-274
    [arXiv:hep-ph/9511244];
%
    Phys.Rev. D {\bf 52} (1995) 2878            %%%-2890.



\end{thebibliography}
%%%%%






% \end{thebibliography}

\end{document}